\documentstyle[aps,multicol,pre,epsfig]{revtex}

\newcommand{\dd}{d}

\newlength{\moreinterligne}
\newlength{\letterarrowvskip}
\newcommand{\vect}{\vec}

\newcommand{\lB}{\ell_B}
\newcommand{\lD}{\ell_D}
\newcommand{\kDR}{\kappa_D R}
\newcommand{\kD}{\kappa_D}
\newcommand{\D}{\displaystyle}

\begin{document}

\title{Interaction between charged anisotropic macromolecules : application to rod-like polyelectrolytes}
\author{David Chapot$^1$, Lyd{\'e}ric Bocquet$^2$ and Emmanuel Trizac$^3$}
\address{$^{1}$Laboratoire de Physique de l'E.N.S. de Lyon, UMR 
CNRS 5672, 46 All{\'e}e d'Italie, \\ 69364 Lyon Cedex, France}
\address{$^{2}$Laboratoire PMCN, UMR CNRS 5586, Universit\'e Lyon 1, 69622 Villeurbanne Cedex, France}
\address{$^3$Laboratoire de Physique Th{\'e}orique, UMR CNRS 8627,
B{\^a}timent 210, Universit{\'e} Paris-Sud,
91405 Orsay Cedex, France}
\date{\today}
\maketitle

\begin{abstract} 
In this paper we propose a framework allowing to compute the effective interactions between two 
anisotropic macromolecules, thereby generalizing the DLVO theory to non spherical finite size
colloids. We show in particular that the effective interaction potential 
remains anisotropic at all distances
and provide an expression for the anisotropy factor. We then apply this framework to the
case of finite rod-like polyelectrolytes. The calculation of the interaction energy requires
the numerical computation of the surface charge profiles,
which result here from a constant surface potential on the rod-like colloids.
However, a simplified analytical description  is proposed, leading to an excellent
agreement with the full numerical solution. Conclusions on the phase properties of 
rod-like colloids are proposed in this context.
\end{abstract}


\begin{multicols}{2}
\narrowtext

\section{Introduction}

The DLVO theory, named after Derjaguin, Landau, Verwey and Overbeek \cite{Verwey}, 
is one of the most influential and still very important description
of charged colloidal suspensions. It has been developped more than fifty years ago
to rationalize the stability of lyophobic colloidal suspensions.
One specific prediction of the DLVO theory is the far-field pair potential
between two spherical colloids of like radii $a$ which, 
within a linearization approximation, takes a Yukawa form :
\begin{equation}
U_{\text{12}}(r)= {Z^2\,e^2\over {4 \pi \epsilon}}\left({\exp[\kappa_D a]\over{1+\kappa_D a}}\right)^2 
{\exp (-\kappa_D r)\over r}, 
\label{V_Yukawa}
\end{equation}
where $Z$ is the valence of the
object, $e$ the elementary charge and $\kappa_D $ denotes the inverse Debye screening length. The latter
is defined in terms
of the micro-ions bulk densities \{$\rho_\alpha$\} (with valencies \{$z_\alpha$\}) as : 
$\kappa_D^2=4 \pi \ell_B \sum_\alpha \rho_\alpha z_\alpha^2$. 
At the level of a dielectric continuum approximation for the solvent with permittivity 
$\epsilon$, 
the Bjerrum length
$\ell_B$ is defined as $\ell _{B}=e^{2}/(4\pi \epsilon
k_{B}T)$, where $k_B T$ the thermal energy: 
$\ell _{B}=7\,$\AA\ for water at room temperature. Note that
the Debye screening factor, $\kappa_D $, does characterize the decay rate of the interaction potential in the far 
field region, providing therefore an experimental measurement
of the screening factor from interaction force measurements (see eg. \cite{Crocker94}).

However, in the colloid world, the spherical shape is not the rule and many macromolecules
are intrinsically very anistropic : rod-like or ribbon-like shapes (DNA molecules, TMV or fd virus, 
V$_2$O$_5$ ribbons, Boehmite rods, etc.) \cite{Fraden89,Purdy03,Guilleaume02,Buining94,Pelletier99}, 
disk-like shapes (eg for clays, as laponite, 
bentonite, etc.)\cite{Mourchid95,Mourchid98,Nicolai01,vanderBeer03}. 
Since the seminal work of Langmuir on bentonite clay particles published 
in 1938 \cite{Langmuir38}, these systems have been the object of considerable attention,
in particular in the context of orientational phase transitions (such as isotropic to nematic I-N, etc.) \cite{Odijk86}. 
From the theoretical side, these transitions were first adressed by Onsager \cite{Onsager49}, who showed
that the nematic phase was stabilized at high density by purely entropic effects. The extension to charged
rods has been reconsidered more recently by Stroobants {\it et al.} \cite{Stroobants86}, showing that
the electrostatic interaction between the polyelectrolytes lead to a twisting effect which enhances the 
concentration at the I-N transition. The picture of Onsager reproduces correctly the experimental results
for highly disymmetric particles, such as TMV or fd viruses \cite{Fraden89,Purdy03}. However 
in many anisotropic systems, a gelation occurs before any I-N transition \cite{Buining94,Mourchid95,Mourchid98}.
According to the DLVO theory, gelation is usually assumed to result from the presence of van der Waals attraction
between the macromolecules, which overcome at high salinity the double layer repulsion. However, the
origin of gelation in many rod- and platelet- like systems remains quite obscure \cite{Buining94,Mourchid98,Bonn99}.
The "gel" denomination is also misleading in some cases since the texture of the "gelled" system 
may be closer to a glassy like phase, in which the orientational and translational 
degrees of freedom are frozen \cite{Bonn99,Trizac02}. The origin of such a glass-like transition is 
still under debate.

In this paper, we shall stay at a  more "microscopic" level and consider the effects of anisotropy on 
the interaction between two macromolecules, much in the spirit of the DLVO approach. One specific 
question we raise is the following. We consider two anisotropic particles, separated by a "large" distance 
({\it i.e.} a distance $r$ larger than their typical dimension $a$). 
Can the electrostatic interaction between these
two individual objects be modelled by the previous DLVO result, {\it i.e.} is the anisotropy lost for
large distances ? This is of course the case in the absence of salt\cite{Jackson}. Does
this result generalize with an electrolyte~?

\begin{center}
\begin{figure}
\epsfig{figure=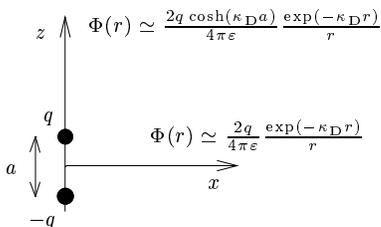}
\caption{Illustration of the anisotropic effect. In the $x$ and $z$ directions, the far-field potentials differ from a factor $\cosh (\kappa_D a)$ which does not vanish at any distance.}
\label{fig:dipole}
\end{figure}
\end{center}

Before delving into the details, let us first consider a much simpler problem, 
namely that of two identical charges $q$, 
with positions $z=\pm a/2$ along the $z$ axis ($a$ fixed) and embeded in an electrolyte (\ref{fig:dipole}): 
what is the electrostatic potential $\phi$ 
created by these two charges at large distances ? 

Naively, one would expect that the anisotropy
is lost for large distances ({\it i.e.} distances larger than the size $a$ of the object,
or larger than the Debye length $1/\kappa_D$) 
and the potential should reduce to its Yukawa form  $\Phi(r)={2q\over {4 \pi \epsilon}}
{\exp (-\kappa_D r)\over r}$. But this is  actually {\bf not} the case ! This can be  understood
by computing --within a linear Debye-H\"uckel like theory-- 
the potential at large distances in the $x$ and $z$ directions : along the axis
$x$, one gets as expected  $\Phi(r)\simeq {2q\over {4 \pi \epsilon}}{\exp (-\kappa_D r)\over r}$ 
to lowest order in $a/r$; but
on the $z$ axis, one gets at the same order $\Phi(r)\simeq {2q \cosh (\kappa_D a) \over {4 \pi \epsilon}}
{\exp (-\kappa_D r)\over r}$. There is consequently a residual anistropy factor (here $\cosh (\kappa_D a)$) between the
two directions, which does not disappear at large distances $r$ from the charges.

The same result is  expected to hold for anisotropic macromolecules, with a residual 
and potentially strong 
anisotropy at large distances. The corresponding
generalization of the DLVO theory is thus required. We emphasize immediatly 
that the proposed description is mostly relevant in the case of moderately dissymetric objects, {\it i.e.}
not too large aspect ratio, since the interaction energy we shall compute is valid for distances 
between the objects larger than their typical size (this precludes infinite objects). This is
anyway the case for many macromolecules (Laponite clays, Boehmite rods, etc.).

The purpose of the present paper is twofold
\begin{itemize}
\item we shall first describe in a general way the far field interaction between two anisotropic 
macromolecules. 
This will lead  to a generalized DLVO interaction between two non-spherical molecules,
with a formal expression of the anisotropic interaction factor. 
\item we shall then apply these results to the case of finite cylinders. 
A byproduct of this part of the work is the charge carried
by the finite cylinder and a description of the edge effects on the cylinders. An approximate
analytical model is proposed yielding results in good agreement with numerical 
calculations. Note that we chose the finite cylinder geometry, not only for its relevance for 
polyelectrolytes, but also because we expect edge effects to be particularly marked. This
geometry is therefore a "benchmark" for the study of anisotropic electrostatic interactions.
\end{itemize}

As in the original calculation of Verwey and Overbeek \cite{Verwey}, the macromolecules
are specified by a constant electrostatic potential on their surfaces and the electrostatic
potential in the electrolyte solution is described at the level of the linearized mean-field
Poisson-Boltzmann equation. However we will show extensively
in a subsequent paper \cite{Chapot2} that this assumption is justified for colloids bearing a 
large constant charge on their surfaces \cite{bta}. For small surface charges, the sketch of resolution 
presented thereafter can also be easily adapted.


This paper is organized as follows:
\begin{itemize}
\item in the first sections, we present the general method we have developped to 
construct the solution of the problem.  
\item we subsequently deduce the general formula for the interaction between two 
anisotropic colloids at large distances. This yields a formal expression of the
anisotropic factors discussed above.
\item we then apply this general method to the specific case of finite cylinders. 
We first obtain the charge distributions on the cylinder, exhibiting the so-called edge
effects. The influence of electrolyte concentration and finite-size effects are discussed.
\item An approximate analytical model is eventually proposed to describe these effects,
yielding results in quantitative agreement with the numerical solution. 
\end{itemize}

\section{General Considerations and description of the problem}

\subsection{Method of resolution : the auxiliary surface charge}

We consider a single charged macromolecule embeded in an infinite electrolyte solution. 
The solution
is characterized by a Debye screening length, $\ell_D=1/\kappa_D$ and
as emphasized above, we assume that the electrostatic potential at the surface of
the macromolecule, $\Phi_0$,  is held constant \cite{Verwey}. 
The electrical double
layer around the macromolecule is described at the level of the linearized Poisson-Boltzmann
theory. This relies on a mean-field description of the micro-ion clouds, 
together with a small potential assumption. An extensive discussion of all these assumptions 
can be found in \cite{Verwey,bta}. We anticipate however that the assumption
of a constant potential at the macromolecule boundary naturally emerges as an 
{\it effective condition} to describe correctly the far field obtained within 
the full
non linear Poisson-Boltzmann theory, for colloids with a large bare charge,
provided $\kappa_D a$ is not too small \cite{bta,Chapot2}.

In this context, outside the macromolecule, the electrostatic potential obeys
the linearized Poisson-Boltzmann (LPB) equation 
\begin{equation}
 \qquad\D\Delta\Phi (\vect{r})=\kappa_D ^2\Phi (\vect{r})  
\label{LPB}
\end{equation}
together with the boundary condition on the molecule surface $\Sigma$
\begin{equation}
 \Phi (\vect{r})=\Phi_0
\label{BC}
\end{equation}
Note that we assume that the macromolecule interior is empty of charges, 
so that $ \Phi (\vect{r})=\Phi_0$ for any point $\vect{r}$ inside the macroion
(this amounts to write $\D\Delta\Phi (\vect{r})=0$).

\begin{center}
\begin{figure}
\epsfig{figure=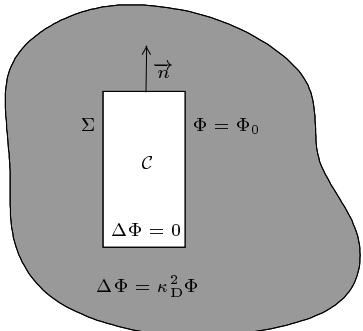}
\caption{Geometry of the problem. A macromolecule ${\cal C}$, with a surface potential $\Phi=\Phi_0$, 
is immersed in an infinite electrolyte. The permittivity of the macromolecule
is assumed to be much lower than that of the solvent(water), 
so that the electrostatic potential is assumed to
be constant in the interior of ${\cal C}$.}
\label{fig:immersed_cyl}
\end{figure}
\end{center}
 
The surface charge density, $\sigma$,  is then obtained from the derivative of the electrostatic 
potential at the molecule surface~:
\begin{equation}
\D \sigma (\vect{r})=-\epsilon \left(\frac{\partial \Phi}{\partial \vect{n}}\right)_{\Sigma^+}
\end{equation}
where $\vect{n}$ is the (outer) unitary vector perpendicular to $\Sigma$
and the notation $\D\left(\frac{\partial \Phi}{\partial\vect{n}}\right)_{\Sigma^+}$ stands
for $\vect{n}\cdot\bbox{\nabla}\Phi$

The standard Green function formalism is too cumbersome to be applied in its simplest version
to solve the previous equations, Eqs. (\ref{LPB}) and (\ref{BC}). This is due to the existence of a 
non-vanishing excluded region for the micro-ions (inside the macromolecule), where the LPB
equation, Eq. (\ref{LPB}), does not apply. In other words, the relevant Green's function
for the problem depends on particule shape and size, which seriously limit its practical
interest.
To circumvent this difficulty, we have therefore introduced
an auxilary system, in which the LPB equation applies everywhere in the volume.
This is defined as~:
\begin{equation}
\left\{\begin{array}{ll}
\D\mbox{for }\vect{r}\in\cal C & \qquad\D\Delta\Phi (\vect{r})=\kappa_D ^2\Phi (\vect{r}) \\
\D\mbox{for }\vect{r}\not\in{\cal C} & \qquad\D\Delta\Phi (\vect{r})=\kappa_D ^2\Phi (\vect{r}) \\
\mbox{for }\vect{r}\in{\Sigma} & \qquad\Phi (\vect{r})=\Phi_0
\end{array}\right.
\label{LPB_bis}
\end{equation}
The corresponding surface charge on the molecule, $\tilde{\sigma}$, is defined here in
terms of the solution $\Phi_{\text{full}}(\vect{r})$ of the previous system of equations~:
\begin{equation} 
\tilde{\sigma} (\vect{r})=\left[\left(\frac{\partial \Phi_{\text{full}}(\vect{r})}{\partial \vect{n}}\right)_{\Sigma^-}-\left(\frac{\partial \Phi_{\text{full}}(\vect{r})}{\partial \vect{n}}\right)_{\Sigma^+}\right]
\label{sigma_tilde}
\end{equation}

Of course, the solution of Eq. (\ref{LPB_bis}), $\Phi_{\text{full}}(\vect{r})$, reduces  to the solution
of Eq. (\ref{LPB}),  $\Phi_{\text{empty}}(\vect{r})$, outside  the macromolecule. This matching
originates in the unicity theorem for the operator $-\Delta +\kappa_D ^2$ with Neumann 
or Dirichlet boundary conditions (see \cite{Jackson} for a similar result concerning
the bare Laplace operator $\Delta$).

Now, the solution of Eq. (\ref{LPB_bis}), $\Phi_{\text{full}}(\vect{r})$ can be defined in terms
of the surface charge $\tilde{\sigma}$~:
\begin{equation}
\Phi_{\text{full}}(\vect{r})=\int\!\!\!\!\int_{\Sigma} \tilde{\sigma} (\vect{r}')\,G(\vect{r},\vect{r}')\,\dd S'
\label{phifull}
\end{equation}
where $G(\vect{r},\vect{r}')$ is the screened electrostatic Green function,
$G(\vect{r},\vect{r}')=\exp(-\kD \vert \vect{r}-\vect{r'}\vert)/(4\pi\epsilon \vert \vect{r}-\vect{r'}\vert)$.
The unknown auxiliary charge, $\tilde{\sigma}$, is found by inverting the boundary condition 
on the macromolecule. This can be explicitly written as: for any point 
$\vec{r}$ on the molecule, 
\begin{equation}
\Phi_0=\int\!\!\!\!\int_{\Sigma} \tilde{\sigma} (\vect{r}')\,G(\vect{r},\vect{r}')\,\dd S'
\label{phiBC}
\end{equation}

The overall result of these general considerations is a formal solution of the LPB equation, 
Eq. (\ref{LPB}), for any point outside the macromolecule:
\begin{equation}
\Phi(\vect{r})=\int\!\!\!\!\int_{\Sigma} \tilde{\sigma} (\vect{r}')\,G(\vect{r},\vect{r}')\,\dd S'
\label{phitot}
\end{equation}
with the auxiliary charge $\tilde{\sigma}$ defined in Eq. (\ref{phiBC}).

To get back to the "real"  charge on the macromolecule, one has to compute
the surface charge density as a function of the auxiliary quantity, $\tilde{\sigma}$.
Using the definition $\sigma (\vect{r})= -\epsilon \left(\frac{\partial \Phi(\vect{r})}{\partial \vect{n}}\right)_{\Sigma^+} $ on any point $\vect{r}$ on the colloid surface $\Sigma$, one obtains:
\begin{equation}
 \sigma (\vect{r})=\,\int\!\!\!\!\int_{\Sigma} \tilde{\sigma} (\vect{r}') \left[ -\frac{\partial \left[G(\vect{r}',\vect{r})\right]}
{\partial \vect{n}} \right]_{\Sigma^+} 
\dd S' 
\label{sigma}
\end{equation}

In practice, the calculation of $\tilde{\sigma}$ which requires the inversion of the boundary 
condition, Eq. (\ref{phiBC}), can be performed analytically  for simple geometries only,
spheres or infinite rods (see below). For more complex case, such as finite
cylinders as considered in this paper, a numerical calculation has to be performed to compute
the inverse matrix of $G(\vec{r}, \vec{r^\prime})$ on the (discretized) macroion. We shall show 
however that a simple model can be proposed which yields results in quantitative agreement with the numerical calculation.

In the case of a given surface charge $\sigma (\vect{r})$, this method of resolution can also be used to calculate the electrostatic potential $\Phi(\vect{r)}$ outside the colloid by computing the auxiliary charge $\tilde{\sigma}(\vect{r}')$ at any point $\vect{r}'$ of $\Sigma$ using Eq. $\ref{sigma}$ and then applying Eq. ($\ref{phitot}$). Once again, the auxiliary charge $\tilde{\sigma}$ is the most relevant parameter to deal with the electrostatic potential created by a colloid immersed in a ionic fluid.

\subsection{The spherical case as an illustrative example}
\label{spherical_case}
Before going further, we come back to the simple spherical problem, where all
previous different quantities, such as the bare and auxiliary surface charge,
can be explicitly computed either by solving the LPB equation straightforwardly, or 
by using the auxiliary charge method sketched above.

We consider an empty sphere $\Sigma$ of radius $a$, at a constant surface potentiel
$\Phi_0$.  On the one hand, the solution of the LPB equation is the usual Yukawa potential~:
\begin{equation}
\Phi (r)=\Phi_0\,a\,\frac{e^{-\kD (r-a)}}{r}
\label{yukawa}
\end{equation}
The surface charge $\sigma$ is defined as $\D\sigma=-\epsilon \frac{\dd \Phi}{\dd r}(r=a)$ and is therefore given by:
\begin{equation}
 \sigma=\epsilon \kD\left(1+\frac{1}{\kD a}\right)\,\Phi_0
\label{sigma_sphere}
\end{equation}

On the other hand, the auxiliary problem described above consists in a sphere $\cal S$ filled with the electrolyte. Using the screened electrostatic Green function
$G(\vect{r},\vect{r}')=\exp(-\kD \vert \vect{r}-\vect{r'}\vert)/(4\pi\epsilon \vert \vect{r}-\vect{r'}\vert)$, 
one may invert the integral equation, Eq. ($\ref{phiBC}$), to obtain the (uniform) auxiliary charge~:
\begin{equation}
\tilde{\sigma}=\epsilon \kD[1+\coth (\kD a)]\,\Phi_0
\label{sigmat_sphere}
\end{equation}
It is then straightforward to show that performing the integral in Eq. ($\ref{sigma}$) allows to recover the
surface charge density obtained above, Eq. ($\ref{sigma_sphere}$). 

This simple example illustrates the difference between the bare and auxiliary problems which we have introduced in the previous section and two ways to calculate the real charge $\sigma$ as a function of $\Phi_0$. The first method could only be used because we knew the formal solution of LPB for a sphere at fixed potential but this is an exception rather than the rule. On the contrary, the auxiliary charge method, even if it seems less sraightforward in this case, is a systematic way to compute the solution of LPB for given boundary conditions.

We now turn to the
calculation of the interaction energy between two macromolecules.

\section{Far-field interaction between anisotropic highly charged colloids}
\label{interaction}

Before focusing on a specific geometry, 
we first use the previous results to describe the interaction
between two anisotropic charged macromolecules.

\begin{center}
\begin{figure}
\epsfig{file=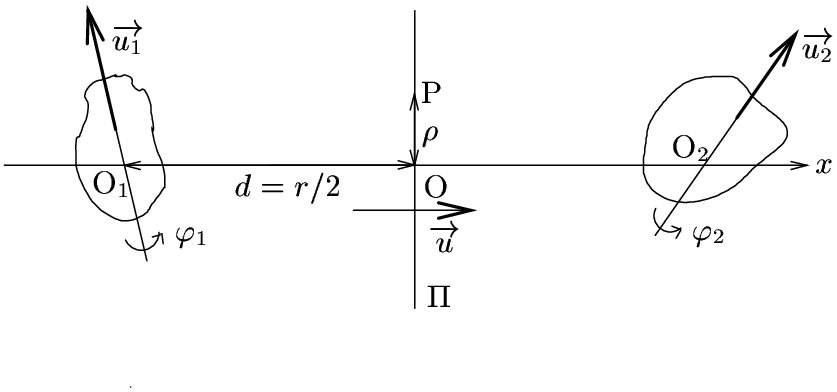,width=8cm}
\caption{Calculation of the electrostatic interaction between two dissymetric macromolecules.
An arbitrary center $O_i$, a unit vector $\vec{u}_i$ and a rotation angle $\varphi_i$ are defined for each molecule. 
We denote as $r=\vert O_1 O_2 \vert$ the distance between the two molecules, while the unit 
vector $\vec{u}$ is defined as $\vec{u}=\vec{O_1 O_2}/\vert O_1 O_2 \vert$. We eventually 
introduce the bisector plane $\Pi$ and the intersection point O between $\Pi$ and $O_1 O_2$.
}
\label{setup}
\end{figure}
\end{center}

We consider two colloids ${\cal C}_i$ ($i=1$, 2) separated by a distance $r$ much larger
than the typical size $D$ of the colloids. As we already noticed in the introduction, it is
important to note that the restriction $r\ge D$ makes sense for moderately dissymetric
macromolecules only.
We assume at this level that the charge profiles $\sigma_i (\vect{r})$, and
equivalently $\tilde{\sigma}_i (\vect{r})$, are known. 
The position of each colloid ${\cal C}_i$ is characterized by fixing a (somewhat arbitrary) origin $O_i$ for the
molecule (this may coincide for example with the colloid center if it is symmetrical). On the other hand, we assume that 
the orientation of the anisotropic colloid is described a unit vector $\vect{u_i}$ pointing into a direction $\Omega_i$ and an angle $\varphi_i$ corresponding to a rotation of ${\cal C}$ around $\vect{u_i}$.
We finally define the colloid-colloid direction using the unit vector $\vect{u}=\vect{O_1 O_2}/\vert O_1 O_2\vert $ 
and introduce the bisector plane, $\Pi$, of $[O_1 O_2]$ and $O$ the intersection of $\Pi$ with $(O_1 O_2)$. It will proove useful to introduce of system of coordinates $\{O,x,y,z\}$,  
with the $x$-axis corresponding to the axis $(O_1 O_2)$ (see Fig. \ref{setup}). The distance
between $O$ and a point $P$ is denoted as $\rho$.

We shall
estimate the  interaction force (acting on one macromolecule due to the other) 
by integrating the electrostatic stress tensor,
$\stackrel{\Rightarrow}{T}$, defined as \cite{Jackson} :
\begin{equation}
\stackrel{\Rightarrow}{T}= \left(P+\frac{\epsilon \vect{E}^2}{2}\right)\stackrel{\Rightarrow}{I}
-{\epsilon} \vect{E}\otimes\vect{E}
\end{equation}
where $\stackrel{\Rightarrow}{I}$ is the identity tensor,
$\vect{E}$ the electrostatic field and $P$ the hydrostatic
pressure.
The force acting on the macromolecule ${\cal C}_2$ can be written accordingly as
\begin{equation}
\D\vect{F}_{2} =\D -\int\!\!\!\!\int_{\Pi} \stackrel{\Rightarrow}{T}\,\vect{dS}
\label{force2}
\end{equation}
Note that the integral runs over the bisector surface $\Pi$, and not the colloid surface. This 
is a consequence of the fact that the divergence of the eletrostatic stress tensor 
$\stackrel{\Rightarrow}{T}$ vanishes outside the macroions.

We emphasize that the following calculations are conducted in the far field
limit where the distance $r$ is larger than the Debye length $\ell_D=\kD^{-1}$. 
This will allow us to expand the various quantities in powers of $\rho/r$. No
specific assumption is done however on the ratio between the typical size of the macromolecule,
$a$, and $\ell_D$. 

Hydrostatic equilibrium and (linearized) Poisson-Boltzmann equations, respectively
$-\vect{\text{grad}}\,p+\rho\,\vect{E}=\vect{0}$ and $\Delta \Phi=\kD ^2 \Phi$, 
allow to write $\D P=P_{\infty}+\frac{\epsilon \kD^2 \,\Phi^2}{2}$. Note that the linearization
of the PB equation is fully justified in the present case since in the far field limit ($r\gg \kappa_D^{-1}$) the
dimensionless electrostatic potential $e\Phi/k_B T$ is expected to be small.
One therefore obtains 
\begin{equation}
\stackrel{\Rightarrow}{T}=\,\left(\,P_{\infty}+\frac{1}{2} \epsilon 
\kD^2 \Phi^2-\frac{\epsilon}{2}\vect{E}^2\right)\,\stackrel{\Rightarrow}{I}-
\epsilon \left(\vect{E}\otimes\vect{E}-\vect{E}^2\right)\,\stackrel{\Rightarrow}{I}
\end{equation}

We denote $E_\alpha$ the component of $\vect{E}$ in the direction $\alpha$,
$\alpha$=$x$, $y$, $z$. Then, for $P\in \Pi$ : $E_y$, $E_z={\cal O}(\rho/r)\,E_x$. Therefore, $\vect{E}^2={E_x}^2\,\left[1+{\cal O}(\rho^2/r^2)\right]$ and $E_\alpha\,E_\beta-{E_\alpha}^2\,\delta_{\alpha \beta}={E_x}^2\,{\cal O}(\rho^2/r^2)$. 

This allows to rewrite the force $\vect{F}_{2}$ acting on the colloid 2 as
\begin{equation}
\D\vect{F}_{2} \simeq \left\{\int\!\!\!\!\int_{\Pi}  \frac{\epsilon}{2}\,\left[\kD^2\Phi^2 (\rho) -{E_x}^2 (\rho) 
\right]\right\}
\vect{dS}
\label{forceapprox}
\end{equation}
Both the potential $\Phi$ and the electric field $E_x$ in this equation 
can be estimated from the solution for the potential created by a single colloid, as 
obtained in the previous paragraph, as we now show. First
Eq. (\ref{phitot}) can be written
\begin{equation}
\Phi(\vect{r})=\int\!\!\!\!\int_{\Sigma} \tilde{\sigma} (\vect{r}') {
\exp(-\kD \vert \vect{r}-\vect{r'}\vert)\over{4\pi\epsilon \vert \vect{r}-\vect{r'}\vert }} dS'
\end{equation}
For distances $r$ much larger than the typical size $a$ of the macromolecule ${\cal C}_i$, 
one might expand the previous equation for small $r'$ to obtain the leading large $r$ contribution:
\begin{equation}
\Phi(\vect{r})={\exp (-\kD r)\over{4\pi \epsilon r}} \int\!\!\!\!\int_{\Sigma} \tilde{\sigma} (\vect{r}') 
\exp(-\kD\vect{u_r}\cdot\vect{r'}) dS'
\end{equation}
with $\vect{u_r}=\vect{r}/r$.
We introduce at this point the total auxialiary charge 
$\tilde{Z}_i=\int\!\!\!\!\int_{\Sigma_i} \tilde{\sigma} (\vect{r}') dS'$ and
the angular distribution
$f_i (P)$ defined as,
\begin{equation}
f_i(P)=1/\tilde{Z}_i \int\!\!\!\!\int_{\Sigma} \tilde{\sigma} (\vect{r}') {
\exp(-\kD \vect{u_r}\cdot\vect{r'})} dS'
\label{angular}
\end{equation}
Using these definitions, one gets eventually the electrostatic potential
at point $P$ as
\begin{equation}
\Phi_i (P)=\frac{\tilde{Z}_i\, \,f_i (P) e^{-\kD r}}{4\pi\epsilon\,r}
\end{equation}
At the order ${\cal O}(\rho/r)$, is is straightforward to check that one might
replace $\vect{u_r}$ by $\vect{u}$ in 
the anisotropic factor $f_i$ of the previous equation : $f_i$ only depends
on the angular coordinates (characterized by $\vect{u_i}$ and $\varphi_i$). Note that the dependance on
$\varphi_i$ disappears for axisymmetric colloids. 
From now on, we will only consider such objects so that may write $f_i=f_i (\vect{u_i})$ for simplification. 
The potential 
created by colloid $i$ therefore reads
\begin{equation}
\Phi_i (P) =\frac{\tilde{Z}_i\, \,f_i (\vect{u_i}) e^{-\kD r}}{4\pi\epsilon\,r}
\end{equation}
In the $r\gg \kD^{-1}$ limit, the corresponding electric field reduces to 
$\vect{E}_i=\pm\,\kD\,\Phi_i (P)\,\vect{u}$, with a plus (resp. minus) sign for $i=1$ (resp. $i=2$).
The total electrostatic potential
$\Phi$ on the mediator plane $\Pi$ is written as the sum of the contributions due to each 
colloids, $\Phi=\Phi_1+\Phi_2$~: 
\begin{equation}
\Phi(P)=\left( \tilde{Z}_1\, \,f_1 (\vect{u_1}) +\tilde{Z}_2\, \,f_2 (\vect{u_2})\right)
\frac{e^{-\kD r}}{4\pi\epsilon\,r}
\end{equation}
Note that the superposition assumption for  the potential is justified
in the far field limit, where one may neglect mutual polarization effects. 
The same holds for the electric field : $E_x=E_1+E_2$,
leading to~:
\begin{equation}
E_x=\kD
\left( \tilde{Z}_1\, \,f_1 (\vect{u_1}) -\tilde{Z}_2\, \,f_2 (\vect{u_2})\right)
\frac{e^{-\kD r}}{4\pi\epsilon\,r} \vect{u}
\end{equation}
Introducing these expressions into Eq. (\ref{forceapprox}) yields the following
expression for the force $\vect{F}_{2}$~:
\begin{eqnarray}
\vect{F}_{2} & = & \D\frac{2\,\kD^2 \tilde{Z}_1\,\tilde{Z}_2\, \,f_1 (\vect{u_1})\,f_2 (\vect{u_2})}{(4\pi)^2\epsilon}\nonumber\\
& \times & \D\int_0^{\infty} 2\pi\,\rho\,d\rho\,\frac{e^{-2\,\kD\sqrt{d^2+\rho^2}}}{d^2+\rho^2}\,\vect{u}
\end{eqnarray}
In the far field region, $r\gg \kD^{-1}$, it is legitimate to expand the integrand in powers
of $\rho/r$ and keep only the leading order : 
using $\D e^{-2\,\kD\sqrt{d^2+\rho^2}}=e^{-\kD\,r\,\left(1+4\,\rho^2/r^2+{\cal O}(\rho^4/r^4)\right)}$, 
one may compute the
integral to get
\begin{equation}
\vect{F}_{ 2} = \frac{\tilde{Z}_1\,\tilde{Z}_2\, \,f_1 (\vect{u_1})\,f_2 (\vect{u_2})\,e^{-\kD r}}{4\pi\epsilon\,r}\,\kD\,\vect{u}
\end{equation}
which is always repulsive \cite{Repuls}.
This force derives form the potential energy (again at leading order in $\kD r$)~:
\begin{equation}
U_{12}(r)=\frac{\tilde{Z}_1\,\tilde{Z}_2\, \,f_1 (\vect{u_1})\,f_2 (\vect{u_2})\,e^{-\kD r}}{4\pi\epsilon\,r}
\label{interaction_energy}
\end{equation}

This expression for the interaction energy between the two macromolecules is one of
the main results of this paper. This generalizes the DLVO calculation for anisotropic 
molecules. Note that, in view of the various expansions performed, this expression 
is valid in the far-field limit, {\it i.e.} for interparticles 
distances $r$ larger than both the Debye length and
the typical size of the colloid $a$ (say, to fix the ideas, $r> 4 \ell_D, 4 a$).

As anticipated in the introduction, the interaction does not reduce at
any distance to the isotropic DLVO result, obtained for spheres. The anisotropy
of the interaction is described by the angular distribution $f_1 (\vect{u_1})$ 
and $f_2 (\vect{u_2})$ defined in Eq. (\ref{angular}). The latter is defined in terms
of the (auxiliary) charge distribution on the macromolecules $\tilde{\sigma} (\vect{r})$,
or equivalently as a function of the bare surface charge $\sigma (\vect{r})$ using
Eq. (\ref{sigma}).

We conclude this part by showing that the previous expression for the interaction
energy indeed reduces to the standard DLVO expression for spheres (as
it should). In this case, the bare and auxiliary surface charge on one sphere
have been computed in the previous section, in Eqs. (\ref{sigma_sphere}) and (\ref{sigmat_sphere}). On the other hand, the angular factor $f_i$ for each
sphere $i$ can be easily computed and reduces to $f_i=\frac{\sinh \kD a}{\kD a}$. 
The latter is of course independent of any angular variable.
Gathering these results, one retrieves the DLVO expression, Eq. (\ref{V_Yukawa})~:
\begin{equation}
U_{12} = {\left(\frac{e^{\kD a}}{1+\kD a}\right)}^2\,\frac{Z_1\,Z_2\,e^{-\kD r}}{4\pi\epsilon\,r}
\end{equation}

A final note concerns the case of colloids with vanishing internal volumes. In the latter
case, the bare and auxiliary charge coincide,  $\tilde{\sigma}=\sigma$, and our calculation 
leads back to the expression found in a different context by Trizac {\it et al.} \cite{Trizac02}:
\begin{equation}
U_{12}=\frac{Z_1\,Z_2\,e^{-\kD r}}{4\pi\epsilon \,r} f_1 (\vect{u_1})\,f_2 (\vect{u_2})
\end{equation}


\section{Charge distribution on a finite rod-like polyelectrolyte}

We now use the previous results to predict the far field interaction between two 
finite rod-like polyelectrolytes.
In contrast to the spherical case, briefly considered in the previous section, the
surface charge  cannot be obtained analytically in this situation.
Therefore, we shall first obtain numerically the surface charge on the
cylinder, by solving Eq. (\ref{phiBC}). We will then
propose a simple analytical model yielding an approximate surface
charge in good agreement with the "exact" numerical results.  

We emphasize at this point that the finite cylinder geometry  
should be merely considered here as a generic situation where end effects are important.
The present description could be easily extended to other related geometries, 
like sphero-cylinders, ellipsoids, etc., though no fondamental difference is however expected.

\subsection{Sketch of the numerical method}
\begin{center}
\begin{figure}[h]
\epsfig{file=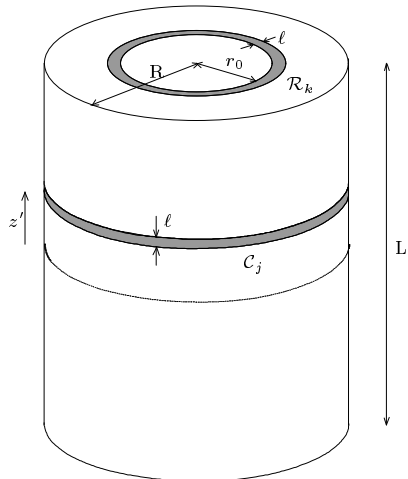}
\caption{The numerical calculations are performed by decomposing the cylinder $\cal C$ into small 
cylinders ${\cal C}_j$ of radius $R$ and height $\ell$, and in rings ${\cal R}_k$ of radii $r_0$ and of width $\ell$.
Each of these elementary surfaces carry a uniform surface charge density $\tilde{\sigma}_i$. 
The numerical calculations were performed with $\ell\leq 0.05\,\lD$.}
\label{fig:cyl_decomposition}
\end{figure}
\end{center}

We now consider a cylinder $\cal C$ with radius $R$ and length $L$, at a contant potential
$\Phi_0$. 
The resolution first starts with the computation of the auxiliary surface charge by inverting
Eq. (\ref{phiBC}). This calculation involves the Green function $G(\vec{r},\vec{r}')$, 
$G(\vect{r},\vect{r}')=\exp(-\kD \vert \vect{r}-\vect{r'}\vert)/(4\pi\epsilon \vert \vect{r}-\vect{r'}\vert)$,
expressing the potential at point $\vec{r}'$ created by a unit point charge in $\vec{r}$.
However due to the cylindrical symetry of the problem, one might reduce the dimensionality
of the problem by integrating the Green function on a ring (or small cylinder) whose center 
matches the axis of symetry of the cylinder, as illustrated on Fig.\ref{fig:cyl_decomposition}. 
This specific problem is considered below.  Once the corresponding reduced
Green function is known, the numerical task simplifies into a standard inversion problem. First,
the cylinder $\cal C$ is decomposed into the superposition of small cylinders 
(on the lateral surface) or rings (on the head surfaces), denoted as ${\cal C}_j$ and ${\cal R}_k$, with dimension $\ell$ and surface charge density 
$\tilde{\sigma}_i$ (see Fig.\ref{fig:cyl_decomposition})).

Then Eq. (\ref{phiBC}) is discretized according to the equation :
\begin{equation}
\forall j\in\Sigma, \Phi (\vect{r_j})=\sum_i\tilde{\sigma}_i\,G_i
(\vect{r_i},\vect{r_j})=\Phi_0
\label{eq:Phi0}
\end{equation}
where $\displaystyle G_i (\vect{r_i},\vect{r_j})$ is the
electrostatic potential created on the cylinder ${\cal C}_j$ or ring ${\cal R}_j$
by the cylinder ${\cal C}_i$ or ring ${\cal R}_i$, carrying a unit
surface charge density. 

\subsection{Reduced Green function}

As mentionned above, the previous inversion requires the knowledge
of the potential created by an elementary ring or cylinder, which we now compute. 
To this end, we make use of the explicit expression of 
the electrostatic potential created by a disk of radius
$R$ at heigh $z'$ carrying a uniform surface charge density (here equal to unity) and
immersed in an electrolyte with Debye length $\lD$. This expression
can be found in Ref.  \cite{WS} and reads ~:
\begin{eqnarray}
G_{\text{disk}}(R,r,z) & = & \D\frac{R}{2\epsilon}\int_{0}^{\infty} \frac{J_1
(k R)\,J_0 (kr)}{\sqrt{k^2+\kD ^2}} \nonumber \\
& \times & \exp\left(-\sqrt{k^2+{\kappa_D}^2}\,|z-z'|\right)\,\dd k
\label{Gdisk}
\end{eqnarray}
with $J_0$ and $J_1$ the Bessel functions of order 0 and 1.
This is namely the potential created by a {\it disk} with radius $R$ at a point M, 
with cylindrical coordinates $\{r,z\}$ (the origin being placed at the center of the disk).
Note also that the dimension of $G_{\text{disk}}$ is given by $R/\epsilon$, since $G_{\text{disk}}$ is the
potential created by a unit surface charge.

Now the potential $dG_{\text{cyl}}(R,r,z)$ created, at a point M, by an infinitesimal {\it cylinder} with height $dz'$, radius R and unit surface charge
can be deduced directly as~:
\begin{equation}
dG_{\text{cyl}}(R,r,z)= dz' {\partial G_{\text{disk}}(R,r,z)\over {\partial R}}
\end{equation}
This leads to 
\begin{eqnarray}
\dd G_{\text{cyl}}(R,r,z) 
& = & \D\frac{R\dd z^{'}}{2\epsilon}\int_{0}^{\infty} \frac{k\,J_0
(kr)\,J_0 (kR)}{\sqrt{k^2+\kD ^2}} \nonumber \\[4mm]
& \times & \D \exp\left(-\sqrt{k^2+{\kappa_D}^2}\,|z-z'|\right)\,\dd k
\label{Gcyl0}
\end{eqnarray}
where the identity $\D \frac{\dd}{\dd x} [x\,J_1 (x)]=x\,J_0 (x)$ has been used.

As a result, the electrostatic potential created by a 
cylinder of radius $R$, height $\ell$ and unit  surface charge, with a center located in $(0,z')$,
is given by
\begin{eqnarray}
G_{\text{cyl}}(R,z',r,z) & = & {R\over 2\epsilon} \D\int_{z'-\ell/2}^{z'+\ell/2} {\dd z''}\int_{0}^{\infty} \frac{k\,J_0
(kr)\,J_0 (k R)}{\sqrt{k^2+\kD ^2}} \nonumber \\
& \times & \D\exp\left(-\sqrt{k^2+{\kappa_D}^2}\,|z-z''|\right)\,\dd k
\label{Gcyl}
\end{eqnarray}

Along the same lines, the potential $\D G_{\text{ring}} (r_0,r,z)$ created by the ring of radius $r_0$ and of thickness $\ell$ can be expressed in terms of $\D G_{\text{disk}} (r_0,r,z)$ according to the relation
\begin{equation}
G_{\text{ring}} (r_0,r,z)=G_{\text{disk}} (r_0+\ell/2,r,z)-G_{\text{disk}} (r_0-\ell/2,r,z)
\label{Gring}
\end{equation}
where $\D G_{\text{disk}} (R,r,z)$ is given above in Eq. (\ref{Gdisk}).

Note that in order to avoid numerical problems, 
the previous integrals must be reformulated specifically for the case $z=z'$. 

\subsection{Calculation of the surface charge}

Inversion of the equation (\ref{eq:Phi0}) yields the auxiliary surface charge $\tilde{\sigma}$.
The "real" surface charge, $\sigma$, can  be deduced from $\tilde{\sigma}$
using Eq. (\ref{sigma}). 
In a discretized form, this reads~:
\begin{equation}
\forall j\in\Sigma, \sigma (\vect{r_j})=\sum_i\tilde{\sigma}_i\,\frac{\partial G_i
(\vect{r_i},\vect{r_j})}{\partial \vect{n_j}}
\label{sigmatilde_sigma}
\end{equation}
where $G_i$ takes either the cylinder or the ring form, obtained in Eqs. (\ref{Gcyl}) and (\ref{Gring}).
This equation involves various derivatives of the Green function at the cylinder surface, namely~:
$\left(\frac{\partial G_{\text{cyl}}}{\partial r}\right)_{r=R^+}$, $\left(\frac{\partial 
G_{\text{cyl}}}{\partial z}\right)$, $\left(\frac{\partial G_{\text{disk}}}{\partial r}\right)_{r=R^+}$
and $\left(\frac{\partial G_{\text{disk}}}{\partial z}\right)$. 

It will turn useful to write all the results 
in terms of dimensionless variables. All the lengths (such as $\ell_B$, $\kD ^{-1}$
or $L$) are expressed in units of the radius of the cylinder $R$~:  eg $L^{\text{adim}}=L/R$. 
In the same way, the electrostatic potential $\Phi$ and surface charge densities $\sigma$ 
become respectively $\displaystyle \Phi^{\text{adim}}=e\Phi/k_B T$ and $\displaystyle 
\sigma^{\text{adim}}=4\pi\lB R\,\sigma/e$ where we recall that $\lB$ is the Bjerrum length defined by 
$\D\lB=e^2/(4\pi\,\varepsilon\,k_B T)$ (for water at room temperature, $\lB=7\,$\AA). 
We also introduce dimensionless Green functions, as $G^{\text{adim}}=\epsilon G/R$
(see previous remark on the dimension of $G$).
From now on, the index ''adim'' will be ommited to simplify notations.

\subsection{Numerical Results}

The previous equations are easily implemented numerically, provided the various
expressions of the Green functions are written in terms of well-converging integrals
as mentionned above.

To fix  ideas the potential on the macromolecule is assumed to be $V_0\simeq100$ mVolts,
so that $\Phi_0=4$ (see however Refs. \cite{bta} and \cite{Chapot2} for further justifications 
of this choice).

\subsubsection{Surface charge profiles} 
\label{surfacechargeprofile}
We now present the results for the surface charges on the lateral and the head of
the cylinder, that we shall denote respectively as $\sigma_{\text{cyl}}(z)$ and 
$\sigma_{\text{head}}(z)$. We first focus  on the shape of
the profiles.
 
\begin{center}
\begin{figure}[h]
\epsfig{file=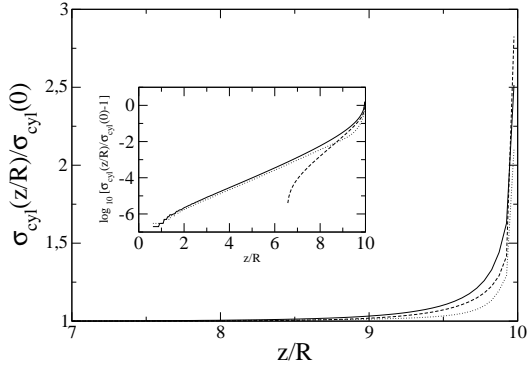,angle=-90}
\caption{Plot of the reduced surface charge on the lateral side of the cylinder, $\ln [\sigma_{\text{cyl}}(z/R)/\sigma_{\text{cyl}}(0)-1]$. 
The aspect ratio of the cylinder is $L/R=20$ and the screening factor is $\kDR=1.0$.Note that $z$ is in units of the cylinder radius $R$.
The solid line is the result of the full numerical calculations, while the dashed line is the
result of the "four parameter" model described in the appendix. The dotted line is the (reduced) auxiliary surface charge  $\tilde{\sigma}_{\text{cyl}}(z/R)/\tilde{\sigma}_{\text{cyl}}(0)$.
Note that the edge effect spans over a smaller distance for the auxiliary surface charge, compared to the "real" charge. See text for details.}
\label{fig:n_sigma_cyl_L_1.0}
\end{figure}
\end{center}

\begin{center}
\begin{figure}[h]
\epsfig{file=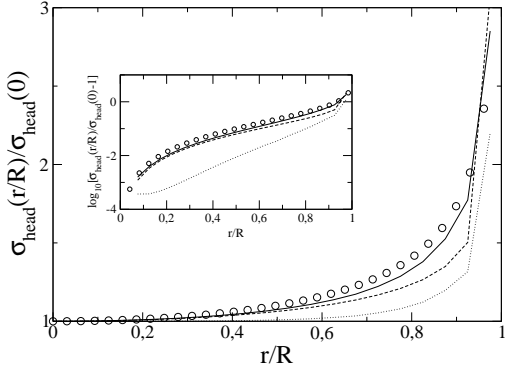,angle=-90}
\caption{Same as Fig. \protect{\ref{fig:n_sigma_cyl_L_1.0}}, but for the surface charge profile 
on the head of the cylinder $\ln [\sigma_{\text{head}}(z/R)/\sigma_{\text{head}}(0)-1]$.
On this figure, we have also plotted the predicted scaling for the divergence in the absence of
salt $\kDR=0$, $\sigma_{\text{head}}(z/R)/\sigma_{\text{head}}(0)=(1-(r/R)^2)^{-1/3}$
(open circles).}
\label{fig:sigma_head_L_1.0}
\end{figure}
\end{center}

Typical results for these profiles are shown on Figs. \ref{fig:n_sigma_cyl_L_1.0}
and \ref{fig:sigma_head_L_1.0}.

Qualitatively, the main striking feature of these profiles is the diverging
surface charge close to the edges of the cylinder. This is of course the well-known
edge effect which is expected for charged objects with uniform potential. In the
absence of electrolyte ($\kD=0$), the divergence of the surface charge in the 
vicinity of an edge is a classical result \cite{Jackson}. For an infinite conducting diedre
with an edge angle  $\beta$, the surface charge density $\sigma$ 
is found to diverge in the 
vicinity of the edge as $\rho^{\pi/\beta-1}$ where $\rho$ is the distance to the edge $\cite{Jackson}$. In the present geometry, corresponding to $\beta=3\pi/2$, the surface
charge is expected to diverge as $\rho^{-1/3}$. For a charged object embedded in 
an electrolyte, {\it i.e.} $\kD\ne 0$, the situation is more complex. However the
divergence is expected to remain, as can be understood from a simple argument. 
As mentionned in paragraph \ref{spherical_case}, the surface charge on a sphere
with radius $a$ and constant potential $\Phi_0$ reads 
$\sigma=\epsilon \kD\left(1+\frac{1}{\kD a}\right)\,\Phi_0$
(see Eq. (\ref{sigma_sphere})). Now using this relationship for a non spherical object, 
one finds that the surface charge $\sigma$ diverge at the points where the radius of curvature
$a$ vanishes. 

Figs. \ref{fig:n_sigma_cyl_L_1.0} and \ref{fig:sigma_head_L_1.0} show that the auxiliary 
surface charge $\tilde{\sigma}$ also exhibits an edge effect. However the latter is more localized close to the edge, 
compared to the "real" surface charge $\sigma$. As for $\sigma$, the divergence of $\tilde{\sigma}$ can be
understood using the results for the sphere, 
$\tilde{\sigma}=\epsilon \kD[1+\coth (\kD a)]\,\Phi_0$, which indeed diverges as the radius of curvature
$a$ vanishes.  However, the transition from a small $a$ region to a large
$a$ region is much more marked for the auxiliary surface charge than for the bare
charge. Indeed from the previous expressions for $\sigma$ and $\tilde{\sigma}$, one gets 
$\sigma(a)=\sigma(a=\infty) + {\cal O}(1/\kD a)$, while $\tilde{\sigma}(a)=\tilde{\sigma}(a=\infty)
+{\cal O} (\exp[-\kD a])$. The large $a$ limit is therefore approached much more quickly for the
auxiliary charge than for the bare charge, which is in agreement with the stronger localization of 
the divergence of the auxiliary charge close to the edge.

We now report in more details on the variations of these density profiles when the 
size of the cylinder $L$ and the screening length $\kD^{-1}$ are varied. 
Generally speaking
the geometry of the problem is characterized by two dimensionless quantities  : the
aspect ratio $L/R$ and the amounts of screening $\kD R$.
Some general trends for the surface charge profiles emerge when these
quantities are varied. 
First, the lateral surface profiles is found to saturate
as the aspect ratio $L/R$ goes to infinity. 
On the other hand, the head profile is found to be barely dependent on the aspect ratio. 
One expects in fact that  the cylinder length $L$ will only play a role when it is smaller or 
equal to the Debye length, $\kD^{-1}$, say $\kD L \le \alpha$,
with $\alpha$ of the order of a few units to fix  ideas. Therefore for a given screening $\kD R$, 
the profile is expected to saturate for aspect ratio larger than $L/R \sim \alpha/(\kD R)$.
This rule of thumb is confirmed when $\kD R$ is varied. In the present  study, we have 
verified this assertion in the interval $\kD R\in[0.1;1]$ (data not shown). 
Typically one finds $\alpha \sim 5$.
Finally it is interesting to compare both profiles with the edge effect divergence predicted in the
$\kD=0$ limit, as argumented above. Only the charge profile on the head is found
to be in semi-quantitative agreement with this scaling, as shown on Fig. 
\ref{fig:sigma_head_L_1.0}. Note that
in order to symetrize the predicted divergence, we compare the head profile
with $\sigma_{\text{head}}(r)/\sigma_{\text{head}}(0)=(1-(r/R)^2)^{-1/3}$. On the other hand
this prediction is found to fail for the cylinder surface charge. 
This is expected since in most of the present calculations, the length $L$ of the cylinder
is larger than the Debye length, so that the $\kD=0$ profile is only a very crude
approximation. On the other hand, the radius of the cylinder
is always smaller than the debye length considered, and for the head, the $\kD=0$ profile 
should be a fair but not so bad approximation for $\kD R \le 1$.

\subsubsection{Total lateral and head charge}

\begin{center}
\begin{figure}[h]
\epsfig{file=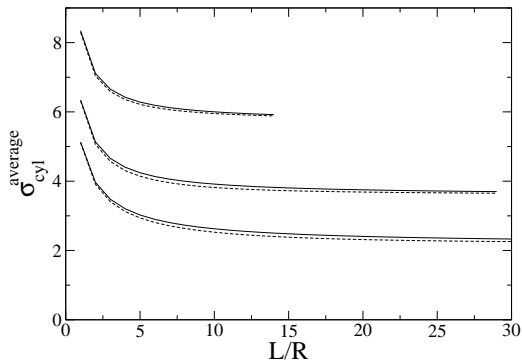,angle=-90}
\caption{Dependance of the averaged surface charge on the cylinder, 
$\D\sigma_{\text{cyl}}^{\text{average}}$, as a function of the aspect ratio 
$L/R$. The solid line is the result of the full numerical calculation, 
while the dashed line corresponds to the four parameter model 
described in the
appendix. From bottom to top, the screening factors are $\kDR=0.2$,
$\kDR=0.5$ and $\kDR=1.0$. The $L=\infty$ asymptotic values are in agreement
with the analytic result, Eq. (\protect{\ref{calculsigmauniforme}}).}
\label{fig:4_parameters_sigma_cyl_average_L_kDR.ps}
\end{figure}
\end{center}

\begin{center}
\begin{figure}[h]
\epsfig{file=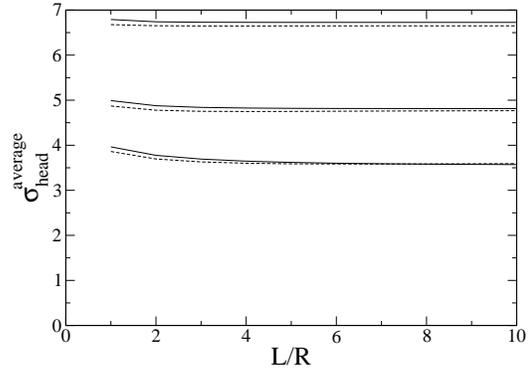,angle=-90}
\caption{Same as  Fig. \ref{fig:4_parameters_sigma_cyl_average_L_kDR.ps}, but with
the charge on the head of the cylinder, 
$\D\sigma_{\text{head}}^{\text{average}}$.}
\label{fig:4_parameters_sigma_head_average_L_kDR.ps}
\end{figure}
\end{center}

A more global quantity of interest is the total charge on the lateral surfaces and on each head of
the cylinder, respectively denoted as $Z_{\text{lat}}e$ and 
$Z_{\text{head}} e$ ($e$ being the elementary charge).  
It prooves in fact useful to consider the average surface charges on the lateral surface
$\sigma_{\text{lat}}^{\text{average}}=Z_{\text{lat}}/(2 \pi R L)$ and on each head of the cylinder,
$\sigma_{\text{head}}^{\text{average}}=Z_{\text{head}}/(\pi R^2)$ (note that we plot below the
{\it reduced} surface charge densities introduced above as $\sigma^*=4\pi \ell_B R \sigma/e$).
These quantities are plotted respectively
on Figs \ref{fig:4_parameters_sigma_cyl_average_L_kDR.ps} and 
\ref{fig:4_parameters_sigma_head_average_L_kDR.ps} as a function of the length
of the cylinder $L/R$ for various screenings $\kD R$.
In the limit of large aspect ratio, both charges saturate to finite values. 
Moreover, both charges are found to be increasing functions of the screening $\kappa_D R$.
This is expected, as can be understood from the spherical test case, 
Eq. (\ref{sigma_sphere}), as a benchmark.

\section{Analytical Description of the surface charge}
\label{analytic}
In this section, we propose a very simplified
description of the electrostatic problem, which has the virtue to provide analytic
estimates of the surface charges. This estimate will proove usefull {\it in fine} to compute
the interaction between two  rod-like polyelectrolytes.
A more detailed approach, including a description of the edge effect, is proposed in appendix
\ref{appE}.

\subsection{Uniform head and lateral surface charges}
\label{uniform_charges}
We consider a "zeroth order" approximation of the problem, consisting in 
a cylinder with uniform charges on the head and on the lateral sides.
More specifically we assume a uniform {\it auxiliary} charge profile. 
We denote $\tilde{\sigma}_{\text{cyl}}$ and $\tilde{\sigma}_{\text{head}}$
the auxiliary surface charge on the cylinder and on the head, and 
by ${\sigma}_{\text{cyl}}$ and ${\sigma}_{\text{head}}$ the corresponding "real" 
surface charges.

At this level of approximation, equation (\ref{eq:Phi0}) relating the {\it auxiliary surface}
charge to the potential $\Phi_0$ reduces to a $2\times 2$ problem~:
\begin{eqnarray}
\Phi_0&=&{\tilde{\sigma}}_{\text{cyl}}\,G_{\text{cyl}}(R,R,0)+ 2\,{\tilde{\sigma}}_{\text{head}}\,G_{\text{head}}(R,R,L/2)\nonumber \\
\Phi_0&=&\D{\tilde{\sigma}}_{\text{cyl}}\,G_{\text{cyl}}(R,0,L/2)\nonumber \\
&+&\D{\tilde{\sigma}}_{\text{head}}\,[G_{\text{head}}(R,0,0)+G_{\text{head}}(R,0,L)]
\label{inversion}
\end{eqnarray}

The surface charges on the head and on the lateral side of the cylinder are then obtained
using Eq. (\ref{sigma}) as
\begin{eqnarray}
\sigma_{\text{cyl}} & = & {\tilde{\sigma}}_{\text{cyl}}\, \left(\frac{\partial G_{\text{cyl}}(R,r,0)}{\partial r}\right)_{R^+}
\nonumber \\
 & + & {\tilde{\sigma}}_{\text{head}}\, \left(\frac{\partial G_{\text{head}}(R,r,L/2)}{\partial r}\right)_{R^+}\nonumber\\
\sigma_{\text{head}} & = & {\tilde{\sigma}}_{\text{cyl}}\, \left(\frac{\partial G_{\text{cyl}}(R,0,z)}{\partial z}\right)_{L/2^+}\nonumber
\\[4mm] & + & {\tilde{\sigma}}_{\text{head}}\, \left(\frac{\partial G_{\text{head}}(R,r,z)}{\partial z}\right)_{0^+}
\label{inversion2}
\end{eqnarray}
In the previous equations, the derivatives of the Green functions are expressed 
in terms of integrals of Bessel functions (see Eqs.  (\ref{Gdisk}), (\ref{Gcyl})), which have to be
computed numerically for any $L$ and $\kappa$.
The systems in Eqs. (\ref{inversion}-\ref{inversion2}) can be easily inverted to obtain the 
expressions of $\D{\sigma}_{\text{cyl}}$ and $\D{\sigma}_{\text{head}}$ as a function of the aspect ratio $L/R$ and screening $\kDR$.

We do not report here the full expressions. Rather we consider the asymptotic $L\rightarrow \infty$
limit, in which the surface charges reach finite values. Note that this limit is reached for
sizes $L$ larger than a few Debye lengths.

In the infinite $L$ limit, the various Green function may be computed, yielding
\begin{eqnarray}
&\D G_{\text{cyl}}(R,R,0)  &= I_0 (\kDR)\,K_0 (\kDR) \nonumber \\
&\D G_{\text{disk}}(R,R,L/2)  &=  0 \nonumber \\
&\D G_{\text{cyl}}(R,0,L/2)  &=  \D{K_0 (\kDR)/2} \nonumber \\
&\D G_{\text{disk}}(R,0,0) &=\D\left(1-e^{-\kDR}\right)/2\,\kDR
\end{eqnarray}

In the same way :
\begin{eqnarray}
&(\partial/{\partial r})&G_{\text{cyl}}(R,r=R^+,0)=\kDR\,I_0(\kDR)\,K_1(\kDR)
\nonumber \\
&(\partial/{\partial z}) &G_{\text{cyl}}(R,0,z=L/2^+)
=\frac{e^{-\kDR}}{2} \nonumber \\
&(\partial/{\partial r}) &G_{\text{disk}}(R,r=R^+,L/2) =  0 \nonumber \\ 
&(\partial/{\partial z}) &G_{\text{disk}}(R,0,z=0^+)
 =\frac{1}{2} 
\end{eqnarray}

Gathering results, we obtain after inversion of Eq. (\ref{inversion}) ~:
\begin{eqnarray}
\D\tilde{\sigma}_{\text{cyl}}^{\text{uniform}} & = & \D\frac{\Phi_0}{I_0 (\kDR)\,K_0 (\kDR)}
\nonumber \\
\D\tilde{\sigma}_{\text{head}}^{\text{uniform}} & = & \D
\frac{2\,\kDR\,\Phi_0}{1-e^{-\kDR}}\,\left[1-\frac{1}{2\,I_0 (\kDR)}\right]
\label{sigmatildeuniform}
\end{eqnarray}
We now denote these
profiles as "uniform" to avoid any confusion with the numerical results.
Using Eq. (\ref{inversion2}), one gets the "real" surface charge densities~:
\begin{eqnarray}
\D\sigma_{\text{cyl}}^{\text{uniform}}& = &\D\Phi_0\frac{\kDR\,K_1 (\kDR)}{K_0 (\kDR)}
\nonumber \\
\D\sigma_{\text{head}}^{\text{uniform}}&=&\D\frac{e^{-\kDR}\,\Phi_0}{2\,I_0 (\kDR)\,K_0 (\kDR)}
\nonumber \\
&+&\D\frac{\kDR\,\Phi_0}{1-e^{-\kDR}}\,\left[1-\frac{1}{2\,I_0 (\kDR)}\right]
\label{calculsigmauniforme}
\end{eqnarray}

In the limit of large $\kDR$, the surface charges are linear in $\kDR$. This is expected
since in this limit, one retrieves the planar results for which $\sigma \propto \kappa \Phi_0$.

The previous result for ${\sigma}_{\text{cyl}}^{\text{uniform}}$ correspond to the semi-infinite cylinder limit \cite{bta}. One may also verify on Fig. 
\ref{fig:4_parameters_sigma_cyl_average_L_kDR.ps} that this result does indeed match 
the $L\rightarrow \infty$ limit of the averaged cylinder profile. 
Note that in contrast, one may verify that the uniform surface charge on the head
$\sigma_{\text{head}}^{\text{uniform}}$ is only a fair approximation to the numerically
computed averaged surface charge, even in the $L\rightarrow \infty$. This is because for the 
screening considered ($\kappa_D R \le 1$), the head always feels the egde of the cylinder.

\subsection{Towards a description of the edge effect}

A simple extension of the previous 
modelization can be proposed : adding a "ring" on the edge of the cylinder should allow to
capture the main features of the edge effect. This can be done in a straightforward way, but
the details of the calculation are somewhat cumbersome. We therefore report the details of this
approach in the apprendix \ref{appE}. This "four parameters" model gives results in good 
agreement with the numerical solutions. This can be seen on Figs. \ref{fig:4_parameters_sigma_cyl_average_L_kDR.ps} and 
\ref{fig:4_parameters_sigma_head_average_L_kDR.ps}, where the results of this
model are displayed (as dashed lines) against the full numerical results.

However,  the interactions between two 
polyelectrolytes do not involve the "real" charge, but the {\it auxiliary charge}. 
As we
show below, the results of the much  simpler "uniform" approach described in 
the previous paragraph will proove sufficient to describe the interaction between 
two rods.

\section{interaction between two rod like polyelectrolytes}

We eventually turn to the description of the interaction between two rod-like polyelectrolytes. 
Our starting point  is the potential energy obtained in  section \ref{interaction}, Eq. (\ref{interaction_energy}).
The two crucial ingredients in this interaction energy are~Ê: the total auxiliary charge $\tilde{Z}$ on the
cylinder;  and the anisotropic term, $f(P)$, defined in terms of the
auxiliary charge profile in Eq. (\ref{angular}). We recall here this expression~: 
\begin{equation}
f(P)=1/\tilde{Z}\int\!\!\!\!\int_{\Sigma} \tilde{\sigma} (\vect{r}') {
\exp(-\kD \vect{u_r}\cdot\vect{r'})} dS'
\label{angular2}
\end{equation}
These ingredients can be therefore easily computed from the full numerical solution, once the
auxiliary surface charge has been computed. 

\subsection{Total auxiliary charge}

\begin{center}
\begin{figure}[h]
\epsfig{file=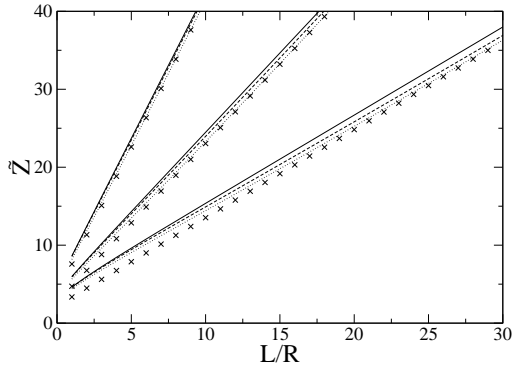,angle=-90}
\caption{Total auxiliary charge $\D\tilde{Z}_{\text{tot}} \ell_B/R$ as a function of the
size of the cylinder $L/R$. The solid line corresponds to the full numerical
resolution, while the crosses are the result of the uniform model. The
dashed line is the result of the four parameter model detailed in the appendix.
The dotted line corresponds to the uniform model with finite $L$ (see text for
details)
From bottom to top the screening factors are $\kDR=0.2$,
$\kDR=0.5$ and $\kDR=1.0$.}
\label{fig:Ztildetot}
\end{figure}
\end{center}

We show on Fig. \ref{fig:Ztildetot} the size dependence of the total auxiliary charge
$\tilde{Z} \ell_B/R$, for various screenings $\kDR$. As can be seen on this
figure, the charge is mainly linear in $L$.
This result is compared with the predictions of the simplified models we have proposed
in the previous section. 
Within the simple uniform surface charge model described in Sec. \ref{uniform_charges},
the total auxiliary charge reads
\begin{equation}
\tilde{Z}=2\pi\,R\,L\,\tilde{\sigma}_{\text{cyl}}^{\text{uniform}}+2\pi\,R^2\,\tilde{\sigma}_{\text{head}}^{\text{uniform}}
\label{zttot}
\end{equation}
Eqs. (\ref{sigmatildeuniform}) reports the expressions of the {\it reduced} auxiliary charges
(recall that in the previous section, reduced variables have been used $\sigma^*=4\pi \ell_B R\sigma /e$).
This leads eventually to the following expression of the total auxiliary charge as a function of the
aspect ratio $L/R$ and screening $\kDR$~: 
\begin{eqnarray}
&\tilde{Z} {\ell_B\over R}=&\Phi_0 \Biggl\{{1\over 2} {L\over R} \frac{1}{I_0 (\kDR)\,K_0 (\kDR)}\nonumber \\
&&+\frac{\kDR}{1-e^{-\kDR}}\,\left[1-\frac{1}{2\,I_0 (\kDR)}\right]\Biggr\}
\label{ZZ}
\end{eqnarray}
This prediction is plotted as crosses on the previous figures, showing a relatively good agreement with
the "exact" numerical results. The agreement might be slightly improved by considering the
complete $L$ dependence, while staying within the uniform model. This corresponds to solving 
the $2\times 2$ system of equations, Eqs. (\ref{inversion2}), with a numerical estimate of the
Green functions for finite $L$. We have plotted the results of this approach as dotted lines on
Fig. \ref{fig:Ztildetot}. This improves slightly the agreement especially for small $L$ and $\kDR$.
We also present the results obtained using the "four parameters" model, described in the appendix.
This model adds to the result in Eq. (\ref{zttot}) the 
contribution of the rings which capture the edge effects.
This model is not analytic either and as can be seen on Fig. (\ref{fig:Ztildetot}), it does not improve
much the agreement.

We conclude here that the very simple analytic expression in Eq. (\ref{zttot}) provides a
useful and trustworthy approximation for the total auxiliary charge which enters the interaction energy,
Eq. (\ref{interaction_energy}).

\subsection{Anisotropic Terms}

\begin{center}
\begin{figure}[h]
\epsfig{file=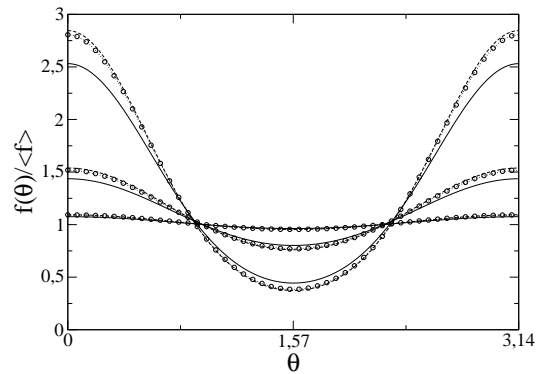,angle=-90}
\caption{Plot of the anisotropic factor of the finite cylinder, $\D f(\theta)/<f>$, as a function 
of the tilt angle. The solid line is the result  of the integration of Eq. (\protect{\ref{angular2}}) over the 
numerically computed surface charge on the cylinder. 
The circles are the result of the uniform model (see text for details) while the
dashed line is the result of the four parameter model described in the appendix. 
The aspect ratio is $L/R=8$ and the screening factors are 
$\kDR=0.2$, $0.5$ and $1.0$ (from bottom to top for $\theta=0$).}\label{fig:ftheta_8}
\end{figure}
\end{center}

\begin{center}
\begin{figure}[h]
\epsfig{file=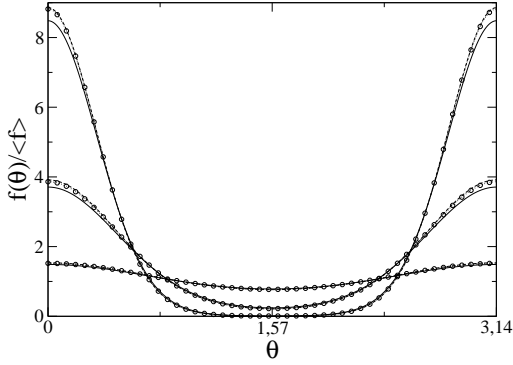,angle=-90}
\caption{Same as Fig. \protect{\ref{fig:ftheta_8}} but for an aspect ratio $L/R=20$}
\label{fig:ftheta_20}
\end{figure}
\end{center}

We report on Figs. \ref{fig:ftheta_8}
and \ref{fig:ftheta_20} the numerical results for the anisotropic terms $f(P)$ for two cylinder sizes $L/R=8$ and
$L/R=20$. These functions have been obtained after numerical integration of Eq. (\ref{angular2}) 
using the numerical result for the auxiliary surface charge.
On these figures, the anisotropic terms are plotted as a function 
of the tilt angle $\theta$, between the axis $z$ of the cylinder considered and the unit vector 
$\vec{u}$ linking the two cylinder centers (see e.g. Fig. \ref{setup}).

It is instructive to compare these "exact" anisotropic factors to the predictions of the simplified models
for the surface charges discussed in the previous section, Sec. \ref{analytic}. Again, let us
first concentrate on the uniform (auxiliary) charge model, proposed in section
\ref{uniform_charges}. In the frame of this simplified description, the anisotropic
factor, in Eq. (\ref{angular2}), can be computed analytically since the auxiliary charges 
are constant over the head and the lateral side of the cylinder. This leads the following 
expression for $f(\theta)$~:
\begin{equation}
 \D f(\vect{n})  = \frac{\tilde{Z}_{\text{cyl}}}{\tilde{Z}}\,f_{\text{cyl}}
(\theta) + 
\frac{\tilde{Z}_{\text{head}}}{\tilde{Z}}\,f_{\text{head}}
(\theta)
\end{equation}

\noindent where $\D\tilde{Z}_{\text{cyl}}={2\pi\,R\,L\,\tilde{\sigma}_{\text{cyl}}}$ is the total charge on the lateral sides
of the polyelectrolyte, and $\D\tilde{Z}_{\text{head}}={2\pi\,R^2\,\tilde{\sigma}_{\text{head}}}$ is the charge on the heads of 
the polyelectrolyte. 
Using expressions, Eqs. (\ref{sigmatildeuniform}) obtained within the uniform model, one has 
\begin{eqnarray}
&\tilde{Z}_{\text{cyl}} \frac{\ell_B}{R}&= {1\over 2} {L\over R}\Phi_0 
\frac{1}{I_0 (\kDR)\,K_0 (\kDR)}\nonumber \\
&\tilde{Z}_{\text{head}}{\ell_B\over R}&= \frac{\kDR\,\Phi_0}{1-e^{-\kDR}}\,\left[1-\frac{1}{2\,I_0 (\kDR)}\right]
\end{eqnarray}
and the total charge $\tilde{Z}$ is given in Eq. (\ref{ZZ}). On the other hand,
the expressions for anistropic factors due to the cylinder and due to the
heads read 
\begin{eqnarray} f_{\text{cyl}} (\theta) & = & \D I_0 (\kD R\,\sin\theta)\,\frac{\sinh\left(\frac{\kD L\,\cos\theta}{2}\right)}{\left(\frac{\kD L\,\cos\theta}{2}\right)}\nonumber\\
f_{\text{head}} (\theta) & = & \D \frac{2\,I_1 (\kDR\,\sin\theta)}{\kD R\,\sin\theta} \cosh\left(\frac{\kD L\cos\theta}{2}\right)
\label{ftheta}
\end{eqnarray}
This expression for $f(\theta)$, using the previous expressions for $\tilde{Z}_{\text{cyl}}$ and
$\tilde{Z}_{\text{head}}$, is plotted
against the numerical results on Fig. \ref{fig:ftheta_8} and  \ref{fig:ftheta_20} for two
aspects ratios ($L/R=8$ and $L/R=20$ respectively). The agreement is seen to be 
surprisingly good in view of the simplicity of the modelization. 

On these figures, we also show the prediction of the more detailed "four parameters" model,
which includes a crude
description of the edge effect, as detailed in appendix \ref{appE}. This approach
adds a contribution from the rings to the previous anisotropic factors, 
$\frac{\tilde{Z}_{\text{ring}}}{\tilde{Z}}\,f_{\text{ring}} (\theta) $
where
$\D\tilde{Z}_{\text{ring}}={2\times 2\pi\,R\,\ell
(\tilde{\sigma}_3+\tilde{\sigma}_4)}$ is the total charge on the rings (see appendix \ref{appE}
for details).  The contribution to the anisotropic factor due to the ring, $f_{\text{ring}}$, 
reads explicitly~:
\begin{equation}
f_{\text{ring}} (\theta)
 =  I_0 (\kD R\,\sin\theta) \cosh\left(\frac{\kD L
\cos\theta}{2}\right)
\end{equation}
As can be seen on the figures, this more detailed description does not improve much the
agreement compared to the much simpler "uniform" approach.

Such a good agreement using a very simple description of the surface charge calls for some 
comments. The crucial point is that the interaction energy involves the {\it auxiliary} charges
and not the bare charges. The full numerical resolution shows in fact that the edge effect is
much more marked for the auxiliary charges than for the "bare" charge, in the sense that the
divergence of the surface charge occurs much closer to the edge for the auxiliary charge.
We have discussed this effect in section \ref{surfacechargeprofile}.
As a result, the auxiliary charge profile is more flat than the "real" charge profile. This feature
allows to understand why the uniform model yields results in good agreement
with the numerical results for the anisotropic factors.

\section{Conclusion}

In the present paper, we have proposed a framework allowing to generalize the DLVO interaction for anistropic macromolecules. 
The central result is the electrostatic interaction energy between two anistropic
macromolecules
\begin{equation}
U_{12}(r)=\frac{\tilde{Z}_1\,\tilde{Z}_2\, \,f_1 (\vect{u_1})\,f_2 (\vect{u_2})\,e^{-\kD
r}}{4\pi\epsilon\,r}.
\label{interaction_energy2}
\end{equation}
The main point resulting from Eq. (\ref{interaction_energy2}) is that 
in a medium with finite salt concentration, the anisotropy is remanent at all  distances.
We have quantified this effect and obtained general formulae for the 
anisotropic factor $f(\vec{u},\varphi)$ (which only depends on 
$\vect{u}$ for axisymmetrical objects) in Eq. (\ref{angular}).
We have then applied this framework to finite rod-like cylinders. 
The previous calculations provide a simple and efficient description of the interaction between two
such polyelectrolytes. In particular, the simple uniform model leads to an analytic expression for
the total auxiliary charge and anisotropic terms which enter the interaction energy, that turn
out to be in good agreement with the full numerical solution.
With this approximation, the anisotropic factor $f(\vect{u})$ for a finite-size 
cylinder of length $L$ and radius $R$ at fixed potential $\Phi_0$ takes a simple form
\begin{equation}
\D f(\vect{u})  = \frac{\tilde{Z}_{\text{cyl}}}{\tilde{Z}}\,f_{\text{cyl}}(\theta) +
\frac{\tilde{Z}_{\text{head}}}{\tilde{Z}}\,f_{\text{head}}(\theta).
\label{f_tot}
\end{equation}
\noindent In the above expression, the auxiliary charges $\tilde{Z}_{\text{cyl}}$, $\tilde{Z}_{\text{head}}$ 
and $\tilde{Z}$, as well as the anisotropy factors  
$f_{\text{cyl}} (\theta)$ and $f_{\text{head}} (\theta)$ are given by 

\begin{eqnarray}
&\tilde{Z}_{\text{cyl}} {\ell_B\over R}&= {1\over 2} {L\over R}\Phi_0 
\frac{1}{I_0 (\kDR)\,K_0 (\kDR)}\nonumber \\
&\tilde{Z}_{\text{head}}{\ell_B\over R}&= \frac{\kDR\,\Phi_0}{1-e^{-\kDR}}\,\left[1-\frac{1}{2\,I_0 (\kDR)}\right]\nonumber\\
&\tilde{Z}&=\tilde{Z}_{\text{cyl}}+\tilde{Z}_{\text{head}}\nonumber
\end{eqnarray}

\noindent and

\begin{eqnarray}f_{\text{cyl}} (\theta) & = & \D I_0 (\kD
R\,\sin\theta)\,\frac{\sinh\left(\frac{\kD
L\,\cos\theta}{2}\right)}{\left(\frac{\kD L\,\cos\theta}{2}\right)}\nonumber\\
f_{\text{head}} (\theta) & = & \D \frac{2\,I_1 (\kD
R\,\sin\theta)}{\kD R\,\sin\theta} \cosh\left(\frac{\kD L
\cos\theta}{2}\right)\nonumber
\end{eqnarray}
As will be shown in $\cite{Chapot2}$, the above expressions with the relevant 
choice of $\Phi_0$ almost corresponds to the interaction energy of two highly 
charged colloids far away from each other, {\em irrespective of their bare charge}.

A few further comments are in order:
\begin{itemize} 
\item First, the interaction energy,  at a {\it fixed center to center
distance between the two cylinders}, is found to be minimum 
when the tilt
angle (made between each cylinder and the center to center direction) is equal to $\pi/2$
i.e. when both cylinders axis are perpendicular to the center to center vector.
Apart from that, the angle between the two axis of the cylinders is
not constrained, at this level of approximation (the two axis may equally
be perpendicular or parallel). This is a consequence of retaining
only the leading order contribution in the potential, and higher order terms
(in $\exp(-\kappa_D r)/r^i$ with $i>1$) would split
the aforementioned degeneracy, and clearly stabilize the crossed rods
compared to the parallel situation.
On the other hand, the interaction is
maximized when the two rods are coaxial (vanishing tilt angle). 
This result somehow contrasts with the infinite rod situation \cite{Stroobants86}, for
which the minimum energy situation corresponds to crossed rods (which is compatible with what we found),
but with a totally different angular dependence, and also a different distance dependence.
\item The anisotropic term in the interaction potential results in a coupling between orientational and translational 
degrees of freedom. The strength of this anisotropy is moerover found to increase with salt concentration.
These ingredients suggest that at high salinity, frustrated phases might form, independently
of van der Waals forces. 
However a full exploration of the phase diagram of charged rods using
these previous results is required before reaching a definite conclusion on the formation
of gels in rod like systems at large salt concentrations, as seen experimentally 
\cite{Buining94,Mourchid95,Mourchid98}. 
\end{itemize}

Work along these lines is in progress.
\vskip 1cm

Acknowledgments: We would like to thank Miguel Aubouy for inspiring discussions
and an enjoyable collaboration on related topics. 


\begin{appendix}

\section{A simple description of the edge effect}
\label{appE}

\subsection{general framework}
\begin{center}
\begin{figure}[h]
\epsfig{file=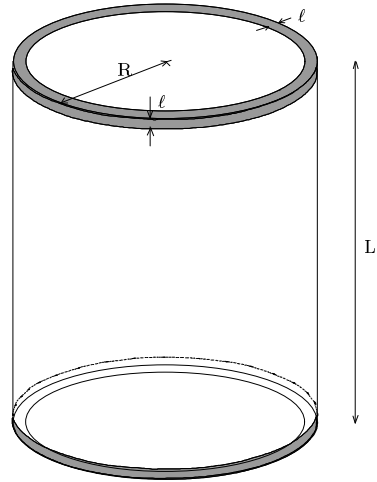}
\caption{Simplified description of the edge effects}
\label{fig:decomposition_4_parameters}
\end{figure}
\end{center}

In this appendix a more detailed description of the edge effect is proposed. 
We extend the model described in section \ref{uniform_charges} by incorporating
a specific charge on the edge of the rod-like macromolecule.
More specifically, we model the {\it auxiliary }surface charge as the superposition
of a uniform charge on the head and on the lateral surface of the cylinder, supplemented by
a ring charge on the edge of the macromolecule, as shown on Fig.
\ref{fig:decomposition_4_parameters}.

From a technical point of view, we separate the ring charge on the edge of the molecule as 
a ring of radius $R$ on the head, and a ring of radius $R$ on the lateral side of the 
cylinder (see figure \ref{fig:decomposition_4_parameters}). The extension of the lateral ring is denoted as $\ell_{\text{cyl}}$, and that
on the edge $\ell_{\text{head}}$.
There is therefore {\it four parameters} in the model: respectively the uniform surface charge on the
head $\tilde{\sigma}_{\text{head}}$, on the lateral sides $\tilde{\sigma}_{\text{cyl}}$,  $\tilde{\sigma}_{\text{cyl. edge}}$ and $\tilde{\sigma}_{\text{head edge}}$.
In the following results, we have chosen $\ell_{\text{cyl}}=\ell_{\text{head}}=0.05R$. Results are only
weakly dependent on this choice. 
As in section \ref{uniform_charges}, one has to solve Eq.
(\ref{eq:Phi0}), relating the auxiliary surface
charge to the potential $\Phi_0$. Within the simplified analysis, and taking into account the symetry
of the cylinder, this equation reduces to a $4\times 4$ inversion~: irrespective of $j$
\begin{equation}
\sum_i\tilde{\sigma}_i\,\tilde{G}_{ij}=\Phi_0
\label{sigmat4}
\end{equation}
where the summation $i$ runs overs the different parts of the simplified object : eg, $i=1$ stands
for the center of the heads of the cylinder; $i=2$ stands for the middle part of the cylinder; while
$i=3$ and $i=4$ stand for the rings on the edges. The "Green functions" $\tilde{G}_{ij}$ are defined
in terms of the Green functions $G_{\text{disk}}$, $G_{\text{cyl}}$ and $G_{ring}$ whose expressions are
given respectively in Eqs. (\ref{Gdisk}), (\ref{Gcyl}) and (\ref{Gring}) (see below for 
the detailed expressions of the $4\times 4 $ matrix $\tilde{G}_{ij}$). Once the auxiliary charges
$\tilde{\sigma}_i$ are known, one obtains the "bare" charge $\sigma$ {\it everywhere} on the cylinder
using Eq. (\ref{sigma}). This can be written formally~:
\begin{equation}
\sigma(\vec{r})=\sum_i \tilde{\sigma_i} {\partial \tilde{G} (\vec{r}, i) \over {\partial \vec{n}}}
\label{sigma4}
\end{equation}
where the notation $\tilde{G} (\vec{r}, i)$ stands for the Green function computed at point $\vec{r}$
due to charge $i$ defined above; $\partial/\partial{\vec{n}}$ denotes the derivative along the normal
to the surface at point $\vec{r}$. 

The previous equation, Eq. (\ref{sigmat4}), is easily inverted and the corresponding surface charges are plotted
on the previous figures, 
Figs \ref{fig:4_parameters_sigma_cyl_average_L_kDR.ps} and  \ref{fig:4_parameters_sigma_head_average_L_kDR.ps}. As shown
on these plots, the approximate description yields results in excellent agreement with the full numerical
resolution for any aspect ratio $L/R$ and screening $\kD R$.


As a consequence, despite its simplicity, the simplified description of the auxiliary charges
contains most of the physics of the edge effect. Also, as shown by the previous
argument, a better agreement is expected for large $\kD R$.

\subsection{technical details}

The cylinder is decomposed into 4 different pieces:

\begin{itemize}
\item a lateral part of length $L$ and radius $R$
\item two disks of radius $R$
\item two lateral rings of radius $R$, of heigths $\ell_{\text{cyl}}$  and of centers located in $\D [0,\pm(L/2\mp\ell_{\text{cyl}}/2)]$
\item two rings of radius $R-\ell_{\text{head}}/2$ and widths $\ell_{\text{head}}$
\end{itemize}

\noindent  respectively denoted 1,2,3 and 4 and carrying the uniform surface charge densities 
$\tilde{\sigma}_{\text{cyl}}$, $\tilde{\sigma}_{\text{head}}$, $\tilde{\sigma}_{\text{cyl edge}}$ 
and $\tilde{\sigma}_{\text{head edge}}$. To simplify the formulation of the equations, we 
respectively call $G_{\text{cyl}}(r_0,r,\ell,z)$ and $G_{\text{ring}}(r_0,\ell_{\text{head}},r,z)$ 
the electrostatic potentials by a cylinder of radius $r_0$ and of height $\ell$ located in $(r,z)$ 
and by a ring of radius $r_0$ and of width $\ell_{\text{head}}$  in $(r,z)$ with the origin of the 
coordinates $(0,0)$ located in the center of the cylinder or of the ring.

In order to find the auxiliary charges on the disks, rings and lateral sides of the cylinder, one has
to solve the $4\times4$ linear problem, obtained from Eq. (\ref{eq:Phi0}) :
$$\forall j\in\{1,2,3,4\},\sum_i\tilde{\sigma}_i\,A_{ij}=\Phi_0$$

The coefficients $\D A_{ij}$ are given in terms of the expressions of $G_{\text{cyl}}$ and 
$G_{\text{disk}}$ given in Eqs. (\ref{Gcyl}) and (\ref{Gring})~:
\begin{eqnarray}
&\D A_{11}& = \D G_{\text{cyl}}(R,R,L,0)\nonumber\\
&\D A_{12}&= \D 2\,G_{\text{disk}}(R,R,L/2)\nonumber\\
&\D A_{13}& = \D 2\,G_{\text{cyl}}(R,R,\ell_{\text{cyl}},L/2-\ell_{\text{cyl}}/2)\\
&\D A_{14}&=2\,G_{\text{ring}}(R-\ell_{\text{head}}/2,R,L/2)\nonumber
\end{eqnarray}

\begin{eqnarray}
&\D A_{21}&=G_{\text{cyl}}(R,0,L,L/2)\nonumber\\
&\D A_{22} & =  G_{\text{disk}}(R,R,0)+G_{\text{disk}}(R,0,L)\nonumber\\
&\D A_{23} & =  G_{\text{cyl}}(R,0,\ell_{\text{cyl}},\ell_{\text{cyl}}/2)\nonumber\\
&  & +G_{\text{cyl}}(R,0,\ell_{\text{cyl}},L-\ell_{\text{cyl}}/2)\\
&\D A_{24} & =  G_{\text{ring}}(R-\ell_{\text{head}}/2,0,0)\nonumber\\
& &+  G_{\text{ring}}(R-\ell_{\text{head}}/2,0,L)\nonumber\\
\end{eqnarray}

\begin{eqnarray}
&\D A_{31} &=G_{\text{cyl}}(R,R,L,L/2-\ell_{\text{cyl}}/2)\nonumber\\
&\D A_{32} & =  G_{\text{disk}}(R,R,\ell_{\text{cyl}}/2)
+  G_{\text{disk}}(R,R,L-\ell_{\text{cyl}}/2)\nonumber\\
&\D A_{33} & =  G_{\text{cyl}}(R,R,\ell_{\text{cyl}}/2,0)
+  G_{\text{cyl}}(R,R,L-\ell_{\text{cyl}})\\
&A_{34} & =  G_{\text{ring}}(R-\ell_{\text{head}}/2,R,\ell_{\text{cyl}}/2)\nonumber\\
& &+  G_{\text{ring}}(R-\ell_{\text{head}}/2,R,L-\ell_{\text{cyl}}/2))\nonumber
\end{eqnarray}

\begin{eqnarray}
&A_{41} &=\D G_{\text{cyl}}(R,R-\ell_{\text{head}}/2,L,L/2)\nonumber\\
&A_{42} & =  G_{\text{disk}}(R,R-\ell_{\text{head}}/2,0)
+  G_{\text{disk}}(R,R-\ell_{\text{head}}/2,L)\nonumber\\
&A_{43} & =  G_{\text{cyl}}(R,R-\ell_{\text{cyl}}/2,\ell_{\text{cyl}}/2)
 +  G_{\text{cyl}}(R,R-\ell_{\text{cyl}}/2,L-\ell_{\text{cyl}}/2)\nonumber\\
&A_{44} & =  G_{\text{disk}}(R-\ell_{\text{head}}/2,R-\ell_{\text{head}}/2,0)\nonumber\\
& &+  G_{\text{disk}}(R-\ell_{\text{head}}/2,R-\ell_{\text{head}}/2,L)
\end{eqnarray}

Once the $\tilde{\sigma}$ have been calculated, we get $\sigma$ using Eq. (\ref{sigmatilde_sigma}),
which reads within the simplified description~:
\begin{equation}
\forall i\in\{1,2,3,4\},\sigma_i (\vect{r}) =\sum_j\tilde{\sigma}_j\,B_{ij}(\vect{r})
\end{equation}
The coefficients $B_{ij}(\vect{r})$ are given by

\begin{eqnarray}
&B_{11}(z) &= \D \left(\frac{\partial G_{\text{cyl}}(R,r,L,z)}{\partial r}\right)_{R^+}\nonumber\\
&B_{12}(z) & = \D \left(\frac{\partial G_{\text{disk}}(R,r,L/2-z)}{\partial r}\right)_{R^+}\nonumber\\
& &+  \D\left(\frac{\partial G_{\text{disk}}(R,r,L/2+z)}{\partial r}\right)_{R^+}\nonumber \\
&B_{13}(z) & =  \D \left(\frac{\partial G_{\text{cyl}}(R,r,\ell_{\text{cyl}},L/2-\ell_{\text{cyl}}/2-z)}{\partial r}\right)_{R^+}\\
& &+  \D \left(\frac{\partial G_{\text{cyl}}(R,r,\ell_{\text{cyl}},L/2-\ell_{\text{cyl}}/2+z)}{\partial r}\right)_{R^+}\nonumber\\
&\D B_{14}(z) & =  \D\left(\frac{\partial G_{\text{ring}}(R-\ell_{\text{head}}/2,r,L/2-z)}{\partial r}\right)_{R^+}\nonumber\\
& &+  \D\left(\frac{\partial G_{\text{ring}}(R-\ell_{\text{head}}/2,r,L/2+z)}{\partial r}\right)_{R^+}\nonumber
\end{eqnarray}

\begin{eqnarray}
&\D B_{21}(r) & = \D\left(\frac{\partial G_{\text{cyl}}(R,r,L,z)}{\partial z}\right)_{L/2^+}\nonumber\\
&\D B_{22}(r) & =  \D\frac{1}{2}+\left(\frac{\partial G_{\text{disk}}(R,r,z)}{\partial z}\right)_{L^+}\nonumber\\
&\D B_{23}(r) & = \D\left(\frac{\partial  G_{\text{cyl}}(R,r,\ell_{\text{cyl}},\ell_{\text{cyl}}/2)}{\partial z}\right)_{\ell_{\text{cyl}}/2^+}\\
& &+  \D \left(\frac{\partial G_{\text{cyl}}(R,r,\ell_{\text{cyl}},z)}{\partial z}\right)_{(L-\ell_{\text{cyl}}/2)^+}\nonumber\\
&\D B_{24}(r) & =  \D \frac{1_{\text{ring}}(r)}{2}+\left(\frac{\partial G_{\text{ring}}(R-\ell_{\text{head}}/2,r,z)}{\partial z}\right)_{(L-\ell_{\text{cyl}}/2)^+}\nonumber
\end{eqnarray}

\noindent with $\D 1_{\text{ring}}(r)=1$  if $R-\ell_{\text{head}}\leq r \leq R$ and 0 otherwise.

\begin{eqnarray}
&\D B_{31} &=\D \left(\frac{\partial G_{\text{cyl}}(R,r,L,L-\ell_{\text{cyl}}/2)}{\partial r}\right)_{R^+}\nonumber\\
&\D B_{32} & =  \D\left(\frac{\partial G_{\text{disk}}(R,r,\ell_{\text{cyl}}/2)}{\partial r}\right)_{R^+}\nonumber \\
&& +  \D\left(\frac{\partial G_{\text{disk}}(R,r,L-\ell_{\text{cyl}}/2)}{\partial r}\right)_{R^+}\nonumber \\
&\D B_{33} & =  \D\left(\frac{\partial G_{\text{cyl}}(R,r,\ell_{\text{cyl}}/2,0)}{\partial r}\right)_{R^+}\\
& &+ \D\left(\frac{\partial
G_{\text{cyl}}(R,r,\ell_{\text{cyl}},L-\ell_{\text{cyl}})}{\partial r}\right)_{R^+}\nonumber \\
&B_{34} & =  \D\left(\frac{\partial G_{\text{ring}}(R-\ell_{\text{head}}/2,r,\ell_{\text{cyl}}/2)}{\partial r}\right)_{R^+}
\nonumber \\
& &+  \D\left(\frac{\partial G_{\text{ring}}(R-\ell_{\text{head}}/2,r,L-\ell_{\text{cyl}}/2)}{\partial r}\right)_{R^+}\nonumber 
\end{eqnarray}

\begin{eqnarray}
&\D B_{41}&=\D \left(\frac{\partial G_{\text{cyl}}(R,R-\ell_{\text{head}}/2,L,z)}{\partial z}\right)_{L/2^+}\nonumber\\
&\D B_{42} & = {1\over 2}+ \left(\frac{\partial G_{\text{disk}}(R,R-\ell_{\text{head}}/2,L)}{\partial z}\right)_{L}\nonumber \\
&\D B_{43} & =  \D\left(\frac{\partial G_{\text{cyl}}(R,R-\ell_{\text{head}}/2,\ell_{\text{cyl}},\ell_{\text{cyl}}/2)}{\partial z}\right)_{\ell_{\text{cyl}}/2^+} \\
& &+ \D\left(\frac{\partial
G_{\text{cyl}}(R,R-\ell_{\text{head}}/2,\ell_{\text{cyl}},L-\ell_{\text{cyl}}/2)}{\partial z}\right)_{(L-\ell_{\text{cyl}}/2)^+}\nonumber\\
&B_{44} & =  \D {1\over 2}+  \D\left(\frac{\partial G_{\text{ring}}(R-\ell_{\text{head}}/2,R-\ell_{\text{head}}/2,L}{\partial z}\right)_{L^+}\nonumber
\end{eqnarray}
\vspace{1cm}

\end{appendix}

\end{multicols}
\end{document}